\documentclass[reprint,superscriptaddress,amsmath,amssymb,aps,prb,longbibliography]{revtex4-2}
\usepackage[utf8]{inputenc}
\usepackage{graphicx}
\usepackage{dcolumn}
\usepackage{bm}
\usepackage[colorlinks,citecolor=blue,linkcolor=blue,urlcolor=blue]{hyperref}
\usepackage[normalem]{ulem}
\usepackage[T1]{fontenc}

\newcommand{\zhang}[1]{{\color{black}{#1}}}
\begin{document}

\title{Exotic $d$-wave \zhang{Cooper Pair} Bose Metal in two dimensions}

\author{Zhangkai Cao}
\thanks{These authors contributed equally.}
\affiliation{School of Science, Harbin Institute of Technology, Shenzhen, 518055, China}

\author{Jiahao Su}
\thanks{These authors contributed equally.}
\affiliation{School of Science, Harbin Institute of Technology, Shenzhen, 518055, China}
\affiliation{Shenzhen Key Laboratory of Advanced Functional Carbon Materials Research and Comprehensive Application, Shenzhen 518055, China.}

\author{Jianyu Li}
\affiliation{School of Science, Harbin Institute of Technology, Shenzhen, 518055, China}
\affiliation{Shenzhen Key Laboratory of Advanced Functional Carbon Materials Research and Comprehensive Application, Shenzhen 518055, China.}
 
\author{Tao Ying}
\affiliation{School of Physics, Harbin Institute of Technology, Harbin 150001, China}

\author{WanSheng Wang}
\email{wangwansheng@nbu.edu.cn}
\affiliation{Department of Physics, Ningbo University, Ningbo 315211, China}

\author{Jin-Hua Sun$^{\ast}$}
\email{sunjinhua@nbu.edu.cn}
\affiliation{Department of Physics, Ningbo University, Ningbo 315211, China}

\author{Ho-Kin Tang}
\email{denghaojian@hit.edu.cn}
\affiliation{School of Science, Harbin Institute of Technology, Shenzhen, 518055, China}
\affiliation{Shenzhen Key Laboratory of Advanced Functional Carbon Materials Research and Comprehensive Application, Shenzhen 518055, China.}

\author{Haiqing Lin}
\affiliation{Institute for Advanced Study in Physics and School of Physics, Zhejiang University, Hangzhou, 310058, China.}

\date{\today}
\begin{abstract}
\zhang{The study of non-Fermi liquids sheds light on unconventional phenomena in condensed matter systems that lie beyond the scope of Landau Fermi liquid theory. One intriguing example is the Bose metal, characterized by an uncondensed bosonic ground state. However, constructing a Bose metal phase in two dimensions (2D) remains a significant challenge.} Utilizing constraint path quantum Monte Carlo and functional renormalization group methods on a fermionic system with spin anisotropy in a 2D lattice, we reveal the emergence of a Cooper pair Bose metal (CPBM) phase in a highly anisotropic regime ($\alpha < 0.30$) with wide range of filling, as proposed in [A. E. Feiguin and M. P. A. Fisher, Phys. Rev. Lett. 103, 025303 (2009)]. Our findings exhibit a visible nonzero momentum Bose surface in the Cooper-pair distribution function, accompanied by a distinct signal of $d_{xy}$ correlation between pairs. Our results highlight that spin-dependent anisotropy in the Fermi surface leads to versatile pairing forms. Platforms such as ultracold atoms in optical lattices and recently proposed altermagnets hold promise for realizing this intriguing phase.
\end{abstract} 
\maketitle

 

\zhang{Significant progress has been made in understanding the novel quantum properties exhibited by strongly correlated electron systems, many of which defy Landau’s Fermi-liquid theory \cite{landau1957theory}. The non-Fermi liquids are closely related to many interesting phenomena in condensed matter physics, including the unconventional superconductivity~\cite{Lee2006-qq,Spalek2022-kf,Keimer2015-jp} and quantum criticality~\cite{Coleman2005-bs}. One interesting phase falling into this category is the Bose liquid \cite{Phillips2003-dd}, where the bosons remain uncondensed, contradicting the usual tendency of bosons to condense into a ground state \cite{einstein1925,Anderson1995-pw}. This reveals new phases of matter that exhibit metal-like behavior without adhering to traditional electron-based Fermi liquid theory. If one presumes the bosons are in form of Cooper pairs, then they are the dominant charge carriers for the electric transport not only in the superconducting but also in the metallic phases, constitute a conducting quantum fluid instead of a superfluid~\cite{Phillips2003-dd,Feiguin2009-nu}.}



\zhang{Microscopically, the Bose liquid is proposed to exist in the hard-core bosons with ring exchange term \cite{Block2011-fj}, as well as the fermionic model with competing spin-dependent Fermi surface anisotropy and on-site attraction \cite{Feiguin2009-nu}.} The existence of a Bose liquid phase could potentially address the obstacle that has hindered the slave-particle gauge theory from explaining the strange metal phase for over 30 years~\cite{Motrunich2007-rb}, as well as potentially explaining the experimental observation in strange metals~\cite{Phillips2022-fu}. However, the concrete numerical evidences of those theoretical proposals are limited to ladder models, i.e., whether the Bose liquid would manifest in two-dimension (2D) remains an open question. Utilizing the density matrix renormalization group (DMRG) and variational Monte Carlo (VMC) methods, Bose liquid is argued to leave fingerprint in the so-called bosonic $J-K$ model \cite{Block2011-fj,Mishmash2011-wn} on multileg ladders. Gluing the non-trivial bosonic degree of freedom with fermionic spinon, physical electrons emerge, governed by the spin-singlet $t-J-K$ model Hamiltonian, revealing a $d$-wave Bose metal (DBM) phase as the quantum ground state in two-leg ladder~\cite{Jiang2013-io}. 


As another line of attacking the problem, Cooper pair Bose metal (CPBM) has been theoretically proposed to exist in a 2D attractive fermionic Hubbard model with anisotropic spin-dependent hopping term~\cite{Feiguin2009-nu,Feiguin2011-ib}. The anisotropic spin-dependent Fermi surface plus attractive interactions leads to an effective model of Cooper pairs with a ring-exchange term, that may allow to realize a paired but non-superfluid Bose metal phase. The Cooper pairs would form a collective state with gapless excitations along a Bose surface but no condensate in momentum space.  Notably, they find the fingerprint of Bose metal phase in the two-leg ladder \cite{Feiguin2011-ib} and the four-leg triangular ladder \cite{Block2011-is}, respectively. However, similar to the bosonic $J-K$ model, the studies are restricted to quasi-1D system due to the method constraint. Subsequently, by using the diagrammatic Monte Carlo method to study on a square lattice at particular filling $n=1.2$~\cite{Gukelberger2014-sf}, they found a density ordered ground state at strong anisotropy and two p-wave triplet pairing phase at intermediate anisotropy, but no signal of CPBM was detected potentially due to the limitation of detecting specific divergence in the method.

In this manuscript, we found the long debating CPBM phase and interesting evolution of density ordered phase in an attractive interacting fermionic system with hopping anisotropy between two species of spins in full 2D lattice, utilizing the constrained path quantum Monte Carlo~(CPQMC) \cite{Zhang1995-hn} and the functional renormalization group~(FRG) \cite{Wetterich1991-gt}. The solution from CPQMC, which addresses the sign-problem by the constraint path, is validated by the high-precision results from FRG. We use the CPQMC result of finite size lattices $L=8,12,16,20$, and we give implementation details of both methods in the Supplementary Materials~\cite{Cao2024-supp}. The uncondensed Cooper pairs in CPBM form a non-superfluid Bose metal, emerging when a spin-dependent anisotropy suppresses the ordinary $s$-wave superfluid ($s$-SF). The dominant $d$-wave nature of Cooper pair correlation is also observed. With varying filling and anisotropy, other phases like charge density wave~(CDW) near half filling and incommensurate density wave~(IDW) at other fillings are also found.

\begin{figure}[b!]
    \centering
    \includegraphics[width=1\linewidth]{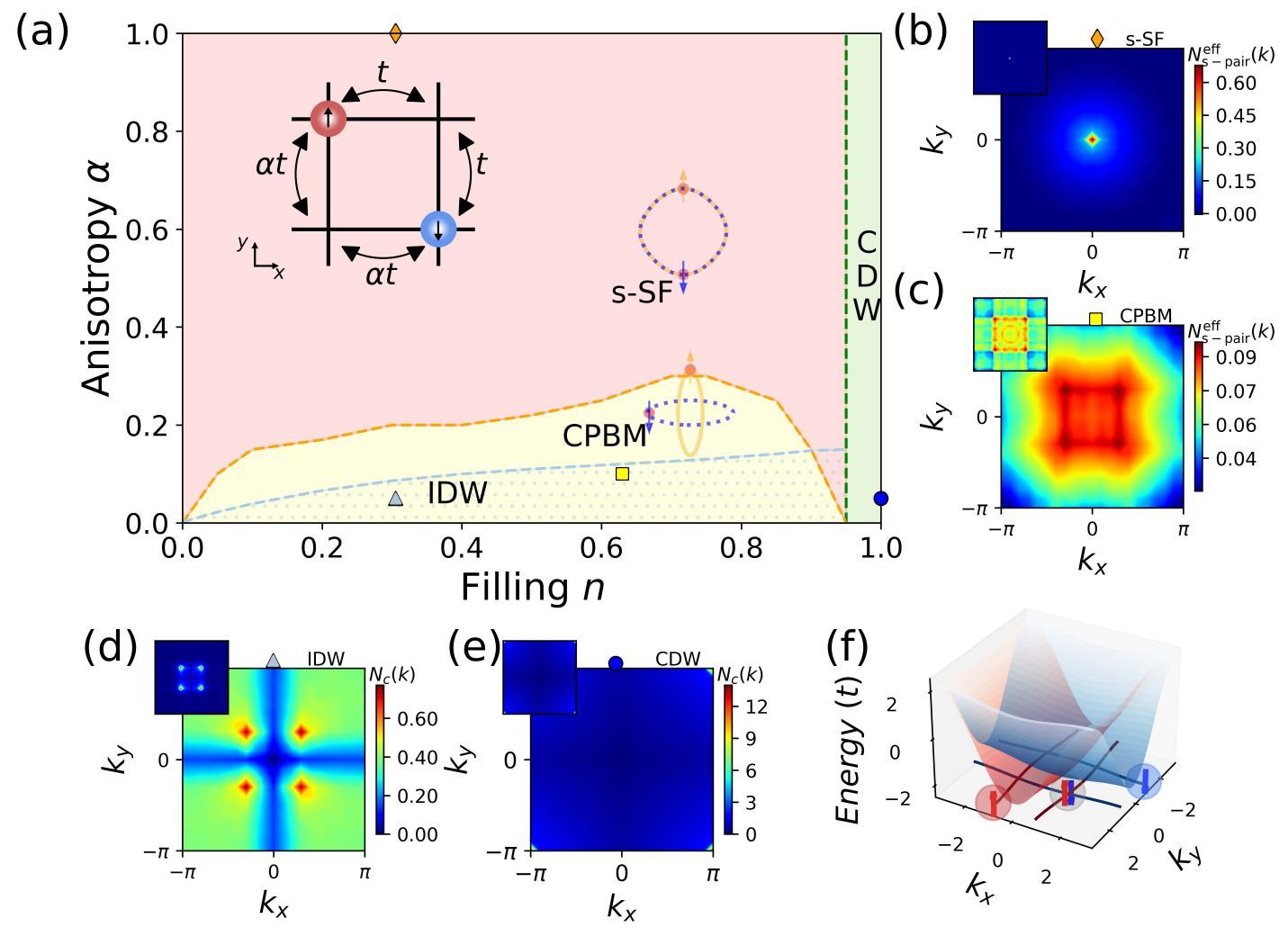}
    \caption{(Color online) \zhang{Model and phase diagram. (a) Schematic zero-temperature phase diagram at $U= -3$ (the unit is $t$), the transition from conventional $s$-wave superfluid ($s$-SF) to  Cooper pair Bose metal (CPBM) is induced by increasing anisotropy. An incommensurate density wave (IDW) at strong anisotropy evolves to a charge density wave (CDW) close to half filling. We shows the effective on-site $s$-wave pair momentum distribution function $N^{\rm eff}_{\mathrm s-pair}({\bf k})$ for (b) s-SF phase and (c) CPBM phase, and the charge structure factor $N_{\mathrm{c}}({\bf k})$ for (d) IDW phase and (e) CDW phase. They show characteristics features in the momentum space. (f) Non-interacting energy spectrum of the spin-dependent Fermi surface with anisotropy $\alpha =0.10$ and filling $n= 0.63$. The qualitative boundary of different phases inferred from our CPQMC data, which we do not claim the preciseness.} }
    \label{fig1}
\end{figure}

The Hamiltonian proposed in the prior studies consists of an anisotropic spin-dependent hopping term~\cite{Feiguin2009-nu}. The term breaks time reversal~(TR) symmetry of the system, which allow lowest energy points shifting from the time-reversal invariant points $(0,0)$ and $(\pi,\pi)$ to other momentum, which is important in forming the CPBM phase. The four-fold rotational ($C_4$) symmetry is broken in the system, however, the Hamiltonian is invariant under combined TR and $C_4$ operation, ensuring that the lowest energy points have fourfold degeneracy. In such a system, the introduction of attraction between fermions could lead to a stable Bose metal phase. Nonetheless, with strong correlation, the numeric is necessary to confirm the exotic phase that potentially triggered by the anisotropy and interaction.

Here, we re-visit the Hamiltonian of Hubbard model on a square lattice with spin-dependent anisotropic hopping given by
\begin{equation}
H =-\sum_{\substack{i, \sigma, l=\hat{x},\hat{y}}} \left( t_{l,\sigma} c_{i,\sigma}^\dag c_{i+l,\sigma} + h.c. \right) + U \sum_i n_{i,\uparrow} n_{i,\downarrow}  
\label{eq:hamiltonian}
\end{equation}
with spin-dependent anisotropic hopping amplitudes $t_{l,\sigma}$, on-site attraction interaction $U < 0$, $c_{i,\sigma}^\dag$ ($c_{i,\sigma}^{\,}$) creation (annihilation) operators with spin $\sigma$ = $\uparrow,\downarrow$, and  $n_{i,\sigma}=c_{i,\sigma}^\dag c_{i,\sigma}$. For defining the anisotropy of fermion hopping, we define a variable $\alpha$, where  $t_{\hat{y}\downarrow} = t_{\hat{x}\uparrow} = t$, $t_{\hat{x}\downarrow} = t_{\hat{y}\uparrow} = \alpha t$ (we assume the nearest-neighbor hopping $t=1$ for simplicity)
leading to an unpolarized system with balanced spin populations, $\langle n_{i,\uparrow} \rangle = \langle n_{i,\downarrow} \rangle = n/2$, we give a schematic diagram on 2D square lattice (see Fig.\ \ref{fig1}(a)).
We defined the anisotropy parameter $\alpha \in [0,1]$, so that $\alpha=1$ corresponds to the isotropic Hubbard model with $C_4$ symmetry and $\alpha=0$ is the extreme anisotropy limit with two-fold ($C_2$) symmetry where fermions can only move in one direction. 

It should be noted that the use of attractive Hubbard $U$ here is a theoretical approach for incorporating the fermionic degrees of freedom within the slave-particle gauge theory~\cite{Feiguin2009-nu,Feiguin2011-ib}, but it does not correspond to the physical parameters of the microscopic model of specific material. When fermions are fully paired in the $\left| U \right| \gg t$ limit, our model can be mapped to an effective boson $J-K$ model: $H_b = -J\sum_{i, j}b_{i}^\dag b_{j} +  K\sum_{ring}b_{1}^\dag b_{2}b_{3}^\dag b_{4} + h.c.$. Here, $b_{i}^\dag = c_{i \uparrow}^\dag c_{i \downarrow}^\dag$, and $i = 1, 2, 3, 4$ labeling sites taken clockwise around a square plaquette. In the effective $J-K$ model, the usual second-order hopping term with strength $J = 2\alpha t^2 / \left| U \right| $, and a fourth-order “ring” exchange term with strength $K = (t^4 + 2\alpha^2t^4 + \alpha^4t^4 )/ \left| U \right|^3$. The detailed discussion of the $J-K$ model is provided in the Supplementary Material~\cite{Cao2024-supp}.  


\zhang{In the context of BCS-BEC theory~\cite{Randeria2012-yd,Chen2024-lu}, a medium or strong attractive Hubbard interaction primarily results in the BEC end of the spectrum. However, the introduction of anisotropy breaks time-reversal symmetry, in which the strong anisotropy leads to a complex uncondensed bosonic phase. The pairing nature between fermions is worth investigating. We defined the on-site $s$-wave pairing operator $\Delta_s^\dagger(i)=c^\dagger_{i\uparrow}c^\dagger_{i\downarrow}$. We measure the effective real-space correlation of the on-site $s$-wave pairing  $C^{\rm eff}_{\mathrm s-pair}(i,j) = \langle {\Delta}_{s}^{\dagger}(i) {\Delta}_{s}(j) \rangle - G^{\uparrow}_{i,j}G^{\downarrow}_{i,j}$, where $G^{\sigma}_{i,j}= \langle c_{i\sigma}c^\dagger_{j\sigma} \rangle$ is the Green's function and $G^{\uparrow}_{i,j}G^{\downarrow}_{i,j}$ is the uncorrelated pairing structure factor.
We measure the effective on-site $s$-wave pairing distribution function in momentum space as the following
\begin{equation*}
N^{\rm eff}_{\mathrm s-pair}({\bf k}) = (1/N)\sum_{i,j} \mbox{exp}[i{\bf k}({\bf r}_i-{\bf r}_j)]C^{\rm eff}_{\mathrm s-pair}(i,j),
\label{nspdtkpair}
\end{equation*}
where $N$ is the number of sites. This is also known as the vertex contribution to pairing. We examined the pairing range of fermions by comparing the strength of on-site s-wave pairing with nearest-neighbor $s$-wave pairing. At the BEC limit, both nearest-neighbor pairing and long-range BCS-type pairing are expected to be suppressed. We found from numerics that under moderate attractive interaction, the pairing distribution function is enhanced in the on-site $s$-wave channel but suppressed in the nearest-neighbor $s$-wave channel. This suggests that the Cooper pairs in our model predominantly exist at the BEC end of the theory, with purely on-site binding forming hard-core bosons. The hopping of on-site paired fermions relates exactly the hopping of boson in $J-K$ model, the spinful fermions contribute either $t$ or $\alpha t$ to the hopping magnitude of the on-site Cooper pair, resulting in the corresponding  $J = 2\alpha t^2 / \left| U \right| $ in the large $U$ limit.}



\zhang{Deduced from $N^{\rm eff}_{\mathrm s-pair}$, we observe a crossover between pairing of conventional s-wave condensation and non-trivial uncondensed Cooper pair Bose metal phase, in the phase diagram as depicted in Fig.\ \ref{fig1}(a). The phase boundaries are estimated using the CPQMC result of the finite size systems, predicting the result at thermodynamics limit~\cite{Cao2024-supp}. We found that the exotic CPBM phase persists from small filling to close to half filling in regions of strong anisotropy, where a spin-dependent anisotropy suppresses the conventional $s$-wave pairing. In regimes characterized by low and medium anisotropy, $s$-SF dominates. A very sharp contrast of $N^{\rm eff}_{\mathrm{s-pair}}({\bf k})$ is found when comparing Fig.\ \ref{fig1}(b) and Fig.\ \ref{fig1}(c), where  the $s$-SF phase exhibits a sharp condensed peak at zero momentum and the CPBM phase exhibits a nonzero momentum continuous uncondensed surface respectively. Another interesting point from the numerics is the obvious transition between the IDW and CDW phases are detected by the charge structure factor, which defined as $N_{\mathrm c}({\bf k}) = (1/N)\sum_{i,j} \mbox{exp}[i{\bf k}({\bf r}_i-{\bf r}_j)]\, \langle {n}_i {n}_j \rangle$ with ${n}_i = \sum_{\sigma} c^\dagger_{i \sigma} c_{i \sigma}$. The IDW phase co-exists with the CPBM phase in extremely strong anisotropy. Fig.\ \ref{fig1}(d) and Fig.\ \ref{fig1}(e) demonstrates the distinguishable features of $N_{\mathrm c}({\bf k})$, in which the sharp peaks are detected in $\mathbf{Q} = (2k_F, 2k_F)$ and $\mathbf{Q} = (\pi, \pi)$ respectively, corresponding to the IDW phase and the CDW phase, where $k_F$ is Fermi momentum defined in the extreme anisotropic limit. In additional to QMC data, FRG provides additional evidence of the CPBM phase and the IDW phase defined in the diagram, showing signals of peaked eigenvalues at nonzero momenta in the $s$-wave pairing channel and the particle-particle channel respectively, as shown in the insets of Fig.\ \ref{fig1}(c) and Fig.\ \ref{fig1}(d). These complimentary evidences consolidate the findings of non-trivial phases.}

\begin{figure}[htb!]
    \centering
    \includegraphics[width=1\linewidth]{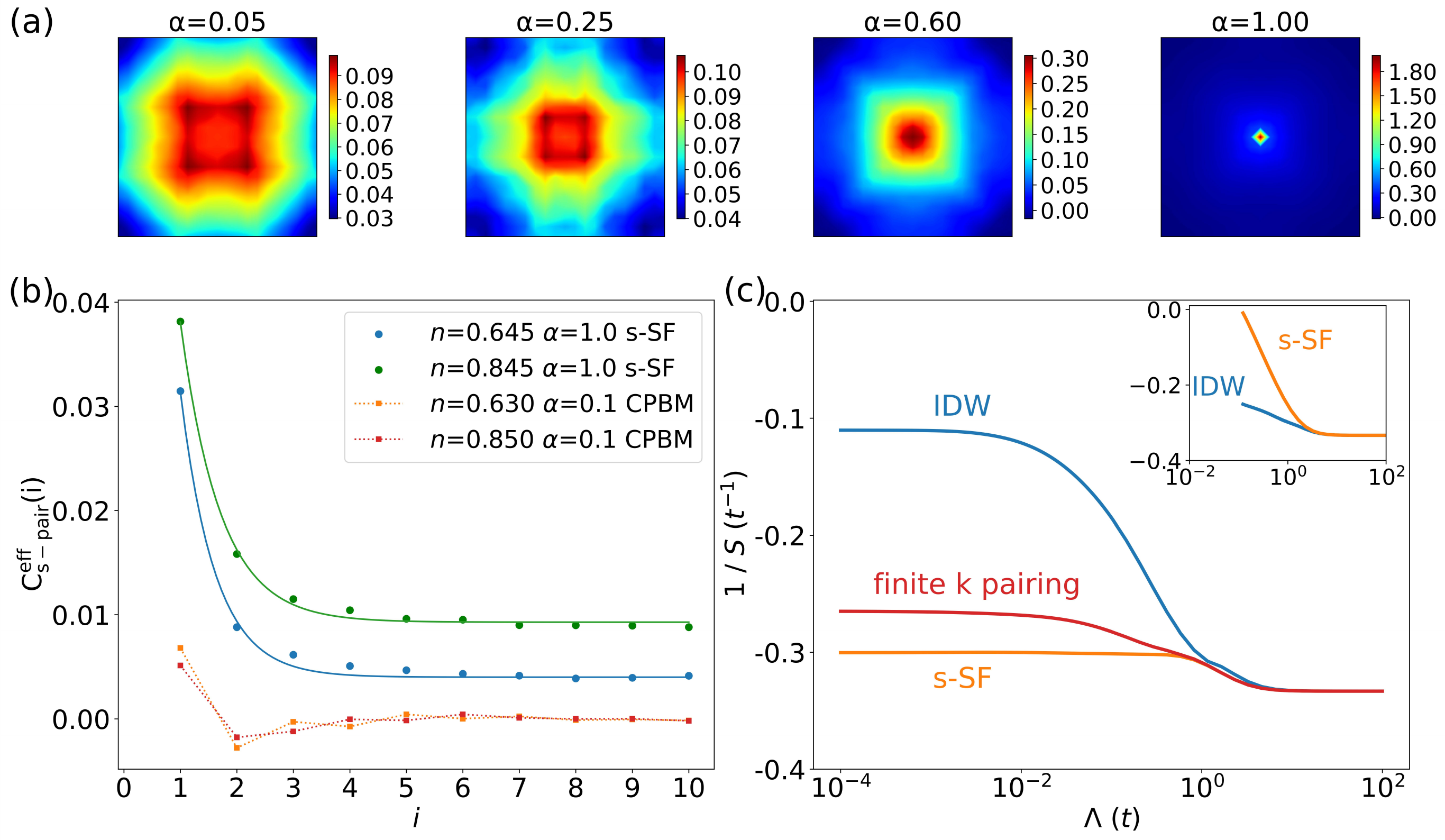}
    \caption{(Color online) Characteristic of CPBM phase. \zhang{(a). CPQMC simulation result of the effective on-site $s$-wave pair momentum distribution function $N^{\rm eff}_{\mathrm s-pair}({\bf k})$ from extremely anisotropy $\alpha \rightarrow 0$ to isotropic $\alpha = 1$ at $n \sim 0.63$ with $U= -3$ on 20 $\times$ 20 lattice, showing the transition of CPBM phase to $s$-SF phase by anisotropy $\alpha$. }(b). The effective real-space correlation of the on-site $s$-wave pairing in the CPBM phase and the $s$-SF phase. The solid line is an exponential fitting curve. (c). The negative leading eigenvalues (NLE) $S$ ($S<0$) with the flow of energy cut off $\Lambda$ in the CPBM phase with $\alpha = 0.10$ at $n = 0.63$. $S$ in all phases are not divergent. The thick red line denotes the nonzero momentum $s$-wave pairing mode, which is the largest pairing modes except for IDW. The inset shows the flow of $S$ in $s$-SF phase with $\alpha =1.0$ at $n= 0.63$, the thick orange line denotes the eventually diverging zero momentum $s$-wave pairing mode.}
    \label{fig2}
\end{figure}

\zhang{We demonstrate the non-interacting energy dispersion for $\alpha =0.10$ in Fig.\ \ref{fig1}(f). The solid red and blue lines below indicate the open anisotropic Fermi surface for a projection at $n = 0.63$. Due to the severe mismatch on the Fermi surface, fermions close to Fermi surface unfavors forming pairs at zero momentum, resulting in pairing tendencies at finite momentum. Pairs with different finite momentum are formed by the fermions at different parts of anisotropic Fermi surfaces, constituting a continuous Bose surface with singular pair distribution function in Fig.\ \ref{fig1}(c), where the weight of pairing is relatively evenly distributed on the Bose surface.} This is the main charisteristic of the CPBM phase, that was proposed as an exotic non-superfluid paired state of fermions~\cite{Feiguin2009-nu,Feiguin2011-ib}. A smaller attraction is insufficient to mediate strong binding of on-site Cooper pairs, resulting in a trivial fermionic metallic state, while a larger attraction results in a conventional superfluid phase.



The size and clarity of the Bose surface are regulated by the filling $n$ and hopping anisotropy $\alpha$. Particularly, in a highly anisotropic regime with $\alpha < 0.30$, Bose surfaces become notably prominent at a filling fraction of $n \sim 0.75$. \zhang{We observed in Fig.~\ref{fig2}(a) that nonzero momentum Bose surfaces only appear in strong anisotropy, and a more pronounced and larger complete Bose surface is formed with increasing anisotropy. When the anisotropy is absent at $\alpha = 1.00$, the system exhibits a strong $s$-SF pairing signal at zero momentum, slowly evolving to weakened $s$-SF with decreasing $\alpha$, finally approaching a small Bose surface formed at $\alpha = 0.25$, and a larger Bose surface in an extremely anisotropic limit.} The difference of the Fermi surface by the two spin electrons increases with the anisotropy, leading to a tendency towards pairing at larger $\bf k$ in the Brillouin zone. Fig.\ \ref{fig2}(b) compares the CPQMC simulation results of the pair correlation function for CPBM phase and $s$-SF phase in real space. The correlation in real space decays to zero quickly in CPBM regime and fluctuates around zero with increasing distance, showing that correlation is short-range. In $s$-SF regime, the correlation shows an exponential decay, and converges to a steady finite value at long distance. This is the signal of the long-range correlation among Cooper pairs.

Fig.\ \ref{fig2}(c) shows the flow of negative leading eigenvalues (NLE) $S$ ($S<0$) in the CPBM phase in FRG. The decrease of the $S$ with the flow of the energy cut off $\Lambda$, indicates the enhancement of fluctuation in the corresponding channel, and the divergence of which signals an emerging order at the associated scattering momentum. The thick red line denotes the nonzero $s$-wave pairing mode, which is the largest pairing modes except for the light blue line which denotes the IDW in particle-hole channel, indicating the nonzero momentum pairing is the highest pairing mode in particle-particle channel. In comparison, the inset shows the flow of $s$-SF phase with $\alpha = 1.0$ at $n = 0.63$, the thick orange line denotes the eventually diverging $s$-wave pairing mode.



\begin{figure}[tb!] 
    \centering
    \includegraphics[width=1\linewidth]{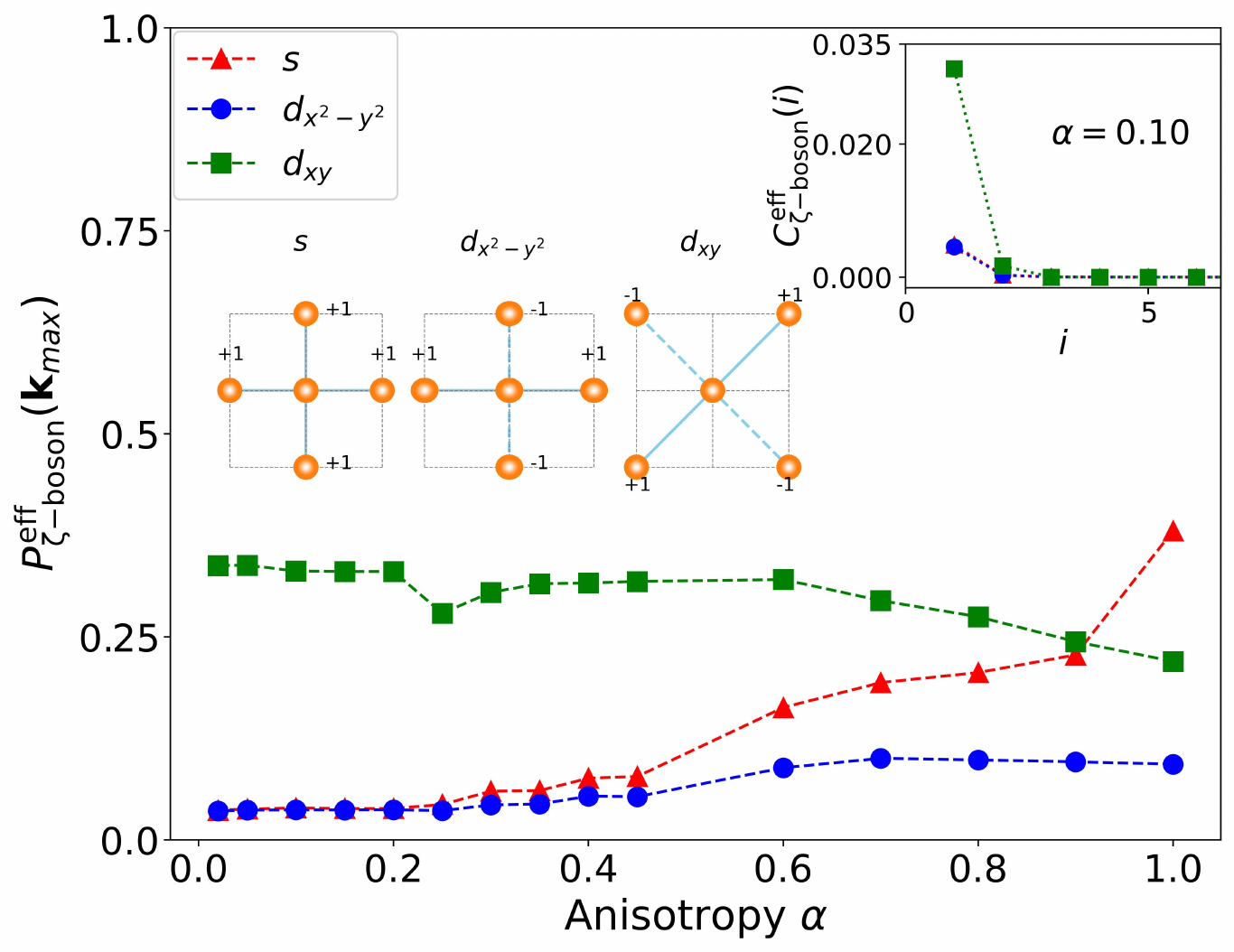}
    \caption{(Color online) Domination of $d_{xy}$-wave boson correlation. The strength of three pairing modes of effective bosons paired correlation at $n\sim 0.63$ with $U= -3$ on 20 $\times$ 20 lattice varies by anisotropy $\alpha$. The inset shows the effective boson pairing correlations in the real space for $\alpha =0.10$. \zhang{We give the schematic of the two-boson (on site pairing) $s$-wave, $d_{x^2-y^2}$-wave, and $d_{xy}$-wave pairing in square lattice.} }
    \label{fig3}
\end{figure}

Another proposed hallmark of the Bose metal phase is its $d$-wave correlation between bosons (on-site pairing), where the $d$-wave here specifically refers to $d_{xy}$-orbital symmetry in consistent with the literature~\cite{Feiguin2011-ib}, exhibiting a propensity to introduce $d$-wave correlations into the system and qualitatively alter the sign structure of the electronic ground state. \zhang{We demonstrated changes in the strength of three pairing modes of boson paired correlation by different anisotropy in Fig.\ \ref{fig3}. To explored the correlation between on-site Cooper pairs $b^\dagger_i=c^\dagger_{i\uparrow}c^\dagger_{i\downarrow}$, we define the effective boson pairing correlations in real space as $C^{\rm eff}_{\mathrm \zeta-boson}(i,j) = \sum_{\delta_{\zeta},\delta'_{\zeta}}(\langle {b}^{\dagger}_{i} {b}^{\dagger}_{i+\delta_{\zeta}} {b}_{j} {b}_{j+\delta'_{\zeta}} \rangle$-$G^{\uparrow}_{i,j}G^{\uparrow}_{i+\delta_{\zeta},j+\delta'_{\zeta}}G^{\downarrow}_{i,j}G^{\downarrow}_{i+\delta_{\zeta},j+\delta'_{\zeta}})$ and the effective two-boson correlator as $P^{\rm eff}_{\mathrm \zeta-boson}({\bf k}) =\sum_{i,j} \mbox{exp}[i{\bf k}({\bf r}_i-{\bf r}_j)]C^{\rm eff}_{\mathrm \zeta-boson}(i,j)/N$. The schematics of different pairing channels $\zeta $ are illustrated, in which $\delta^{(')}_{\zeta}$ are nearest-neighbor sites in the $s$-wave and $d_{x^2-y^2}$-wave pairing, and next-nearest-neighbor sites in the $d_{xy}$-wave pairing.}


The primary competition arises between $s$-wave and $d_{xy}$-wave, while $d_{x^2-y^2}$-wave holds relatively less significance with a small value. As the anisotropy $\alpha$ increases, the system tends toward $d_{xy}$-wave correlation between Cooper pairs. This is the reason why this exotic phase of boson-paired states can be termed the $d$-wave boson-paired state or $d$-wave boson liquid \cite{Sheng2008-jd,Feiguin2011-ib}. In the inset of Fig.\ \ref{fig3}, we compare the correlations between different pairing modes of boson pairs in real space for $\alpha =0.10$. They all decay rapidly to zero in the CPBM regime, indicating short-range correlation among pairs. $d_{xy}$-wave exhibits the highest magnitude, suggesting its predominant boson pairing mode. Our results suggest that the correlation between Cooper pairs is predominantly $d_{xy}$-wave in the CPBM regime with large anisotropy. \zhang{The underlying mechanism of the correlation between pairing could be clearer, when we switch to the picture of $J-K$ model, in which the fourth-order four-site “ring” exchange term with strength $K = (t^4 + 2\alpha^2t^4 + \alpha^4t^4 )/ \left| U \right|^3 $ relates to the hopping of two on-site Cooper pairs on opposite corners of an elementary square plaquette rotating by $\pm 90$ degrees. In the extreme anisotropic limit with $\alpha\rightarrow 0$, the on-site Cooper pair cannot move with $J\rightarrow 0$, leaving $K$ nonzero. So the ring exchange term is increasingly important in the case of large anisotropy, resulting in the dominant $d_{xy}$-wave boson correlation.}

While the extreme anisotropy contributes to the formation of the exotic CPBM phase, it also induce the non-trivial incommensurate density wave in the particle-hole channel. The rich phase diagram results from the interplay between fermionic filling and Fermi surface anisotropy. In this context, the $J-K$ model corresponds to only some aspects of our system. Specific density $n= 1.2$ has been studied \cite{Gukelberger2014-sf} reveal a rich phase diagram, apart from the $s$-SF and IDW phase, notably identifying two different $p$-wave triplet superfluid states with different symmetries at intermediate anisotropy. Additionally, the 2D attractive Hubbard model is known to be dominated by $s$-SF in the absence of anisotropy \cite{Paiva2004-ld}. In our study, we did not find any regions dominated by $p$-wave in the moderately anisotropic region. As the strength of anisotropy increases, the $s$-SF naturally transforms into the CPBM phase in particle-particle channel.

\begin{figure}[htb!]
    \centering
    \includegraphics[width=1\linewidth]{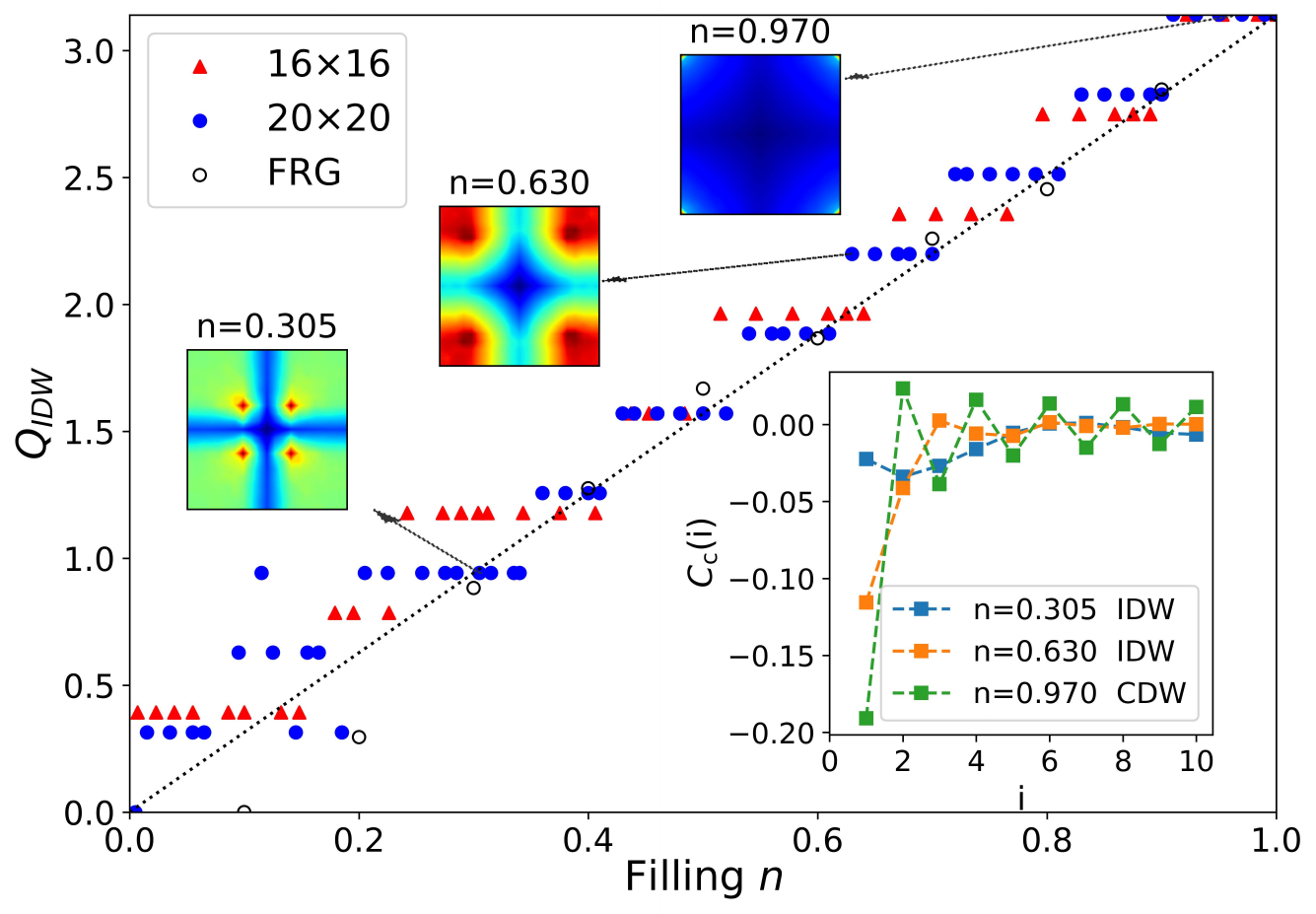}
    \caption{(Color online) Charge density wave and incommensurate density wave. The charge structure factor for $\alpha =0.05$ with different $n$ at $U= -3$. As $n$ increasing, IDW condenses at non zero momentum point $\bf Q$ = ($2k_F$, $2k_F$), gradually diffusing to point $\bf Q$ = ($\pi$, $\pi$) convert to CDW, $\bf Q$$_{IDW}$ is the condensation point in the Brillouin zone. The upper insets show the charge structure factor of the representative IDW to CDW at $n = 0.305$, 0.630, 0.970 on 20 $\times$ 20 lattice, respectively. The bottom inset gives the correlation of charge density wave $C_{\rm c}$ in real space, representing IDW and CDW phase and there is no significant difference from IDW to CDW. In particular, CDW have 2a (lattice constant) periodic modulation of electron density in real space.}
    \label{fig4}
\end{figure}

We identify a very strong commensurate CDW state in particle-hole channel condensed at $\bf Q$ = ($\pi$, $\pi$). While deviating from half-filling, the system exhibits condensation at varying momenta for arbitrary fillings at strong anisotropy, and having singular peaks at the nonzero momentum points.
Fig.\ \ref{fig4} shows IDW condensed at $\bf Q$ = ($2k_F$, $2k_F$). \zhang{The density correlation is expected to oscillate with $2k_F$ wave vector due to the nested condition of the Fermi surface. The peaks gradually diffuse to $\bf Q$ = ($\pi$, $\pi$) as $n$ increases to the half-filling, indicating strong signal of CDW.} Both CPQMC simulation and FRG results unanimously support this inference. The upper insets of Fig.\ \ref{fig4} show the charge structure factor of the representative IDW to CDW at $n$ = 0.305, 0.630, 0.970 respectively. We also define the correlation function of charge density wave in real space $C_{\rm c} = \langle {n}_i {n}_j \rangle$. We show IDW and CDW have algebraically decaying density correlation in real space in bottom inset of Fig.\ \ref{fig4}, and it is long-range order of periodic modulation of electron density.


The CPBM phase can be realized in optical lattice experiments through tuning the filling and hopping anisotropy of effective spin interactions with light. In these experiments one can tune the interparticle interactions to arbitrary values using a Feshbach resonance of tunable few-fermion systems \cite{Serwane2011-yj}, particularly using laser-dressed Rydberg atoms experiments with spin-dependent hyperfine states can reveal their highly tunable range and anisotropy~\cite{Jau2015-kx,Zeiher2016-ls}. Although few-body systems with dipole-dipole interactions have not been studied experimentally as of yet, some theoretical studies of a small ensemble of dipolar bosons are already under investigation \cite{Chatterjee2019-ro,Chatterjee2018-ur}, which can observe $d$-wave correlation between bosons. A recent research observes and quantifies the pseudogap in unitary Fermi gases of lithium-6 atoms through momentum-resolved microwave spectroscopy~\cite{Li2024-cw}. They lend support for the role of preformed pairing as a precursor to superfluidity, offering an ideal quantum simulator to study the proposed Bose metal phase.

More recently, a robust anomalous metallic state with charge-2$e$ quantum oscillations was observed in thin films of $\rm YBa_2Cu_3O_{7-x}$ \cite{Yang2019-is}, potentially indicating a Bose metal phase. 
\zhang{Subsequently, anomalous bosonic metal phases have been observed in diverse systems \cite{Liu2020-mc,Xing2021-xi,Zhang2022-yj,Liao2021-fy,Li2024-us}, still lacks a comprehensive theoretical framework that explains its robustness and experimental properties \cite{Kapitulnik2019-wr,Wang2023-cg}.} The theory of Bose metal may explain this novel phenomenon. 
Lastly, the investigation of spin-dependent anisotropic Fermi surfaces has also been extensive, particularly in the context of altermagnetism~\cite{Smejkal2022-wp,Smejkal2022-po}—a spin-splitting phenomenon with collinear magnetic order incompatible with conventional ferromagnetism or antiferromagnetism. 
The spin-species anisotropy has been found and well studied in recently discovered exotic altermagnets \cite{Smejkal2022-wp,Smejkal2022-po}, exisiting in the materials like $\rm RuO_2$, $\rm KRu_4O_8$, and $\rm La_2CuO_4$. \zhang{These materials might be able to play the role of the fermions with spin-dependent anisotropy in our model, but the microscopic source of the Hubbard attraction still waits to be identified. }
We acknowledge useful discussion with Noah F. Q. Yuan, Xiaosen Yang and Ji Liu.
This work is supported by National Natural Science Foundation of China~(No.~12204130, No.~12088101), National Key Research and Development Program of China (No.~2022YFA1402701). H.K.T acknowledges supports from Shenzhen Start-Up Research Funds~(No.~HA11409065) and Shenzhen Key Laboratory of Advanced Functional Carbon Materials Research and Comprehensive Application~(No.~ZDSYS20220527171407017). J.-H.S. acknowledges supports from the Ningbo Natural Science Fund~(No.~2023J131). T.Y. acknowledges supports from Natural Science Foundation of Heilongjiang Province~(No.~YQ2023A004).

\bibliography{ref}
\end{document}


\title{Supplemental Material for "Exotic d-wave \zhang{Cooper Pair} Bose Metal in two dimensions"}

\author{Zhangkai Cao}
\thanks{These authors contributed equally}
\affiliation{School of Science, Harbin Institute of Technology, Shenzhen, 518055, China}

\author{Jiahao Su}
\thanks{These authors contributed equally}
\affiliation{School of Science, Harbin Institute of Technology, Shenzhen, 518055, China}
\affiliation{Shenzhen Key Laboratory of Advanced Functional Carbon Materials Research and Comprehensive Application, Shenzhen 518055, China.}

\author{Jianyu Li}
\affiliation{School of Science, Harbin Institute of Technology, Shenzhen, 518055, China}
\affiliation{Shenzhen Key Laboratory of Advanced Functional Carbon Materials Research and Comprehensive Application, Shenzhen 518055, China.}
 
\author{Tao Ying}
\affiliation{School of Physics, Harbin Institute of Technology, Harbin 150001, China}

\author{WanSheng Wang}
\email{wangwansheng@nbu.edu.cn}
\affiliation{Department of Physics, Ningbo University, Ningbo 315211, China}

\author{Jin-Hua Sun$^{\ast}$}
\email{sunjinhua@nbu.edu.cn}
\affiliation{Department of Physics, Ningbo University, Ningbo 315211, China}

\author{Ho-Kin Tang}
\email{denghaojian@hit.edu.cn}
\affiliation{School of Science, Harbin Institute of Technology, Shenzhen, 518055, China}
\affiliation{Shenzhen Key Laboratory of Advanced Functional Carbon Materials Research and Comprehensive Application, Shenzhen 518055, China.}

\author{Haiqing Lin}
\affiliation{Institute for Advanced Study in Physics and School of Physics, Zhejiang University, Hangzhou, 310058, China.}

\date{\today} 
\maketitle 

In this Supplemental Material, we present additional computational results. \zhang{In Sec.~S1, we review the background of the model we studied and state the connection between our model of fermions with spin-dependent hopping terms to the $J-K$ model. In Sec.~S2, we discuss the scheme to analyze the data from the finite size system simulation, and determine the phase boundary between Cooper pair Bose metal phase and s-wave superfluid phase, as well as the incommsurate density wave phase and the charge density wave phase. In Sec.~S3, we discuss the background of the numerical methods used, including constraint path quantum Monte Carlo and the functional renormalization group. We discuss the implementation of CPQMC, including the formalism, hyper-parameters and parallelization that is quite necessary for large lattice simulation.}
 
\section{Background of the model of fermions with spin-dependent hopping}
\begin{figure*} 
    \centering
    \includegraphics[width=0.5\linewidth]{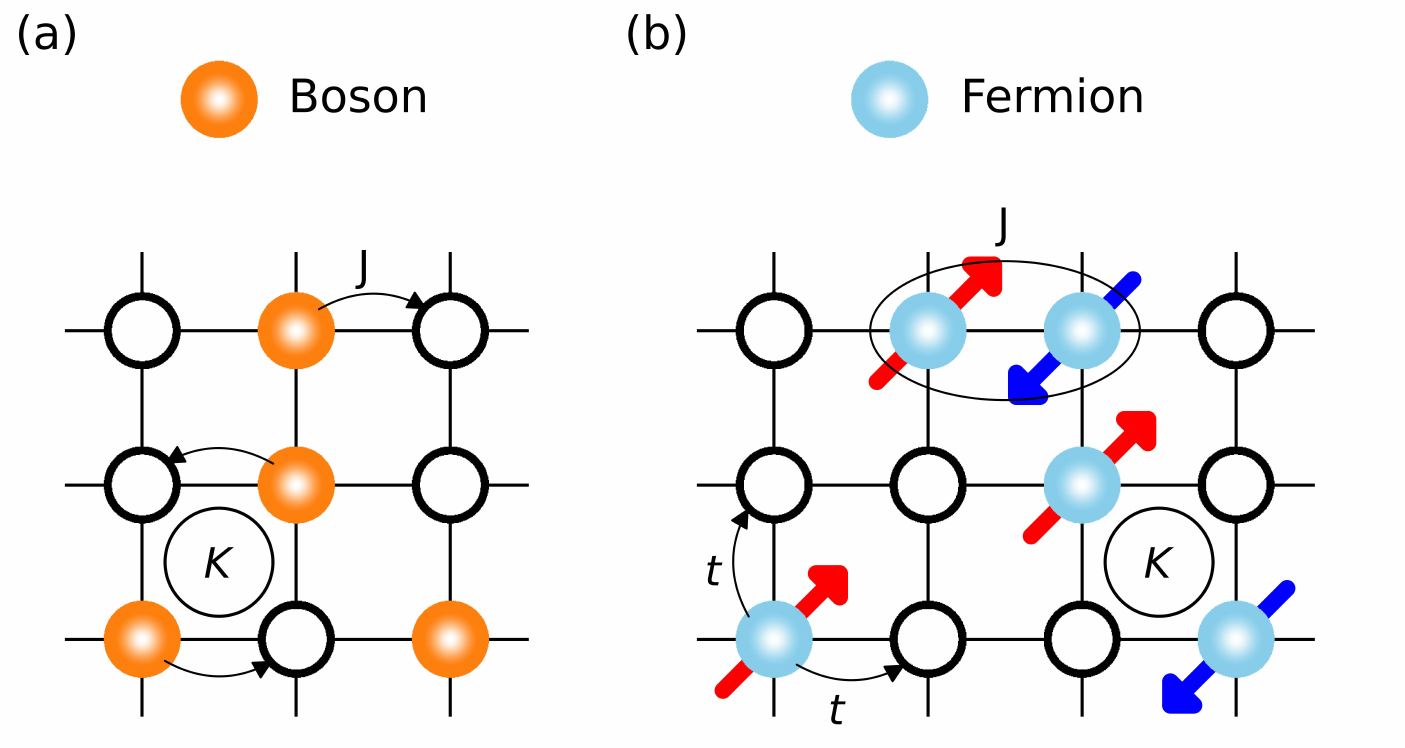}
    \caption{(Color online) \zhang{Schematic of the (a). $J–K$ model, (b). $t–J–K$ model. }}
    \label{figS10}
\end{figure*}

The connection between our model of spin-dependent fermions to the $J-K$ model is made as the following. 
When $\left| U \right| \gg t$, where consider all of the fermions are tightly bound into on-site Cooper pairs. We can derive an effective boson Hamiltonian by considering a perturbation expansion in powers of $t/\left| U \right| $ \cite{Motrunich2007-rb,Feiguin2009-nu,Feiguin2011-ib}. 
In subspaces containing only empty occupancy and paired double occupancy, the effective Hamiltonian can be written as: $H_b = -J\sum_{i, j}b_{i}^\dag b_{j} +  K\sum_{ring}b_{1}^\dag b_{2}b_{3}^\dag b_{4} + h.c.$. Here, $b_{i}^\dag = c_{i  \uparrow}^\dag c_{i \downarrow}^\dag$, and $i = 1, 2, 3, 4$ labeling sites taken clockwise around a square plaquette. \zhang{We show a schematic of the $J–K$ model in Fig.\ \ref{figS10}(a).} In addition to the usual second-order near neighbor boson hopping term with strength $J = 2\alpha t^2 / \left| U \right| $, one obtains a more important fourth-order four-site “ring” exchange term with  strength $K = (t^4 + 2\alpha^2t^4 + \alpha^4t^4 )/ \left| U \right|^3 $, so called $J-K$ model \cite{Feiguin2009-nu,Feiguin2011-ib}. Remarkably, while the hopping term in the extreme anisotropic limit, $\alpha \rightarrow 0$, so $J \rightarrow 0$, but the ring exchange term $K$ which hops pairs of bosons is nonzero. Thus, with increasing anisotropy, we can obtain the ratio $K/J = \frac{1 + 2\alpha^2 + \alpha^4}{\alpha}t^2/\left| U \right|^2$ increases, and the ring exchange term becomes increasingly important in the total Hamiltonian $H_b$. In other words, the magnitude of $K$ is regulated by anisotropy $\alpha$ in our model.

In $t-J-K$ model~\cite{Jiang2013-io}, we first write the electron operator for a bosonic chargon $b(\bf r)$ and fermionic spinon $f_s(\bf r)$, so exactly a chargon and one spinon to compose an electron. Then, we decompose the hard-core boson as $b({\bf r})=d_1({\bf r})d_2({\bf r})$, where $d_1$ and $d_2$ are fermionic slave particles (partons) with anisotropic hopping patterns,  $d_1$($d_2$) is chosen to hop preferentially in the $x(y)$ direction. The electron can written an all-fermionic decomposition format: $c({\bf r})=d_1({\bf r})d_2({\bf r})f_s({\bf r})$. The resulting theory now includes two gauge fields: one to glue together $d_1$ and $d_2$ to form the chargon and another to glue together $b$ and $f$ to form the electron, the first gauge field combine two anisotropic partons to a bosonic chargon, which is the focus of our attention.  The assumption of a positive value for $K$ is crucial in the model to realize the $d$-wave metal. More ab-intio calculations on realistic cuprate could give a reliable estimate of the sign of $K$ as well as its magnitude. \zhang{We show a schematic of the $t-J–K$ model in Fig.\ \ref{figS10}(b).}
Within a gauge theory framework in $J–K$ model, the picture of the Bose metal phase consists of two independent species of fermions hopping on the square lattice with anisotropic hopping, the attractive potential $U$ provides an attractive pairing for bosons. The four-site ring exchange term $K = (t^4 + 2\alpha^2t^4 + \alpha^4t^4 )/ \left| U \right|^3 >0$ in $J-K$ model are four-spin (eight-fermion) terms and can arise from the attractive $U$. 
In contrast to $t-J-K$ model, the electron ring term $K$ are four-fermion terms, which move two charges from one diagonal of a square to the other previously unoccupied diagonal. Thus the electron ring term $K$ can arise from the long-range Coulomb interaction. The strength of the electron ring term increases with increasing hopping anisotropy between the $d_1$ and $d_2$ partons. The attraction here originates from the gauge interaction between the fermionic partons, resulting in bosonic chargon in the slave-particle gauge theory.

\begin{figure*}[]
    \centering
    \includegraphics[width=0.8\linewidth]{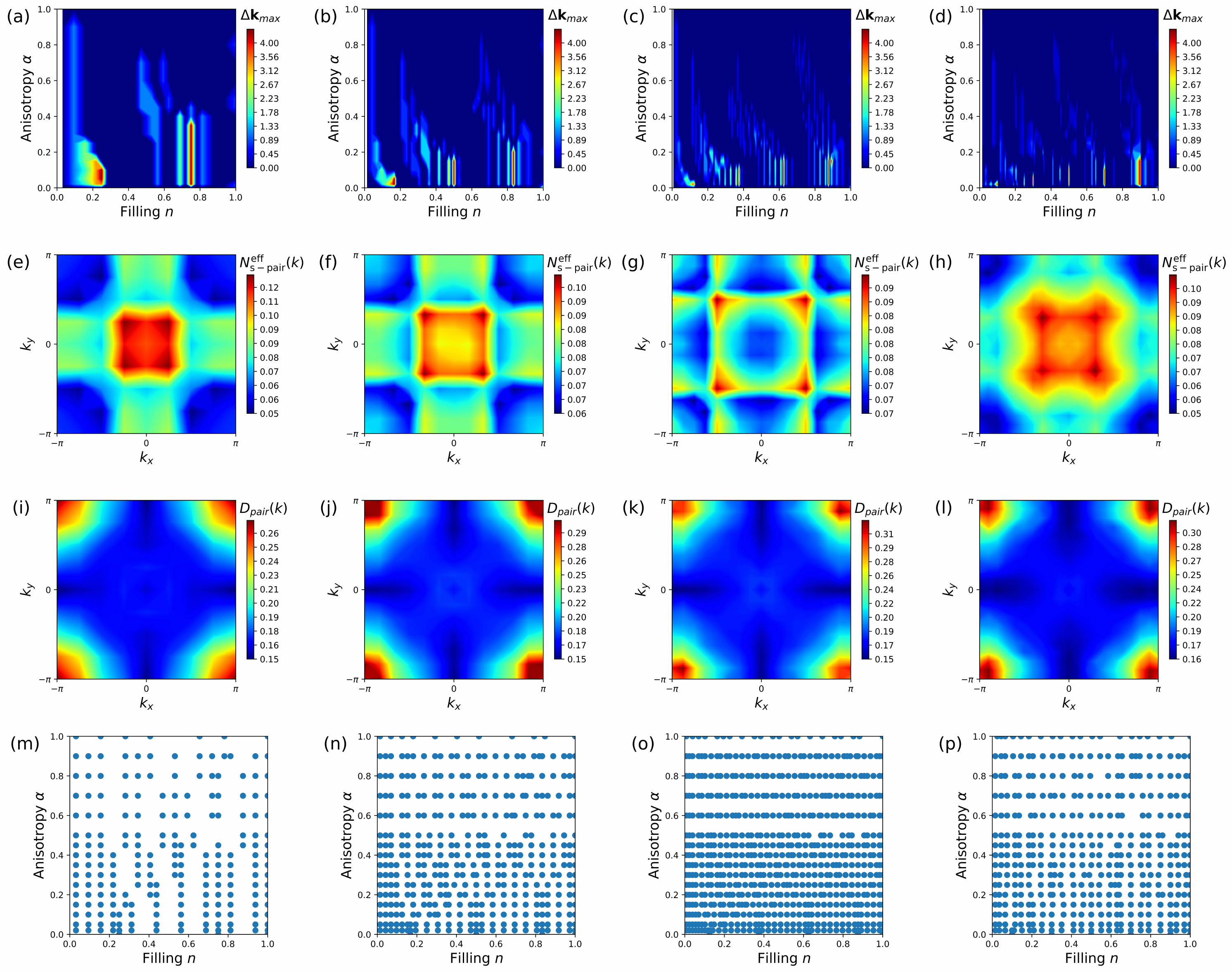}
    \caption{(Color online) Determination of the CPBM. (a-d). CPQMC simulation result for determine the region of CPBM at $U= -3$ on 8 $\times$ 8, 12 $\times$ 12, 16 $\times$ 16, 20 $\times$ 20 lattice, where $\Delta {\bf k}_{\rm max}=\sqrt{{\bf k}_{\rm max}^2(x)+{\bf k}_{\rm max}^2(y)}$, ${\bf k}_{\rm max}(x)$, ${\bf k}_{\rm max}(y)$ are the coordinates of the maximum value in the pair momentum distribution function. The blue areas are $\Delta {\bf k}_{\rm max} = 0$, representing the $s$-wave pair momentum distribution function condensed at $\bf Q$ = (0, 0), other regions are $\Delta {\bf k}_{\rm max} \neq 0$. If $\Delta {\bf k}_{\rm max} \neq 0$, it indicates that there is a nonzero momentum peak in the pair momentum distribution function, and then combined with image display to form a Bose surface, so we believe that there exist a CPBM phase under this parameter. (e-h). CPQMC simulation result of $s$-wave pair momentum distribution function $N^{\rm eff}_{\mathrm s-pair}({\bf k})$ for $\alpha =0.10$ at $n$ $\sim$ 0.85 in $U= -3$ on 8 $\times$ 8, 12 $\times$ 12, 16 $\times$ 16, 20 $\times$ 20 lattice, showing obviously existence of nonzero momentum Bose surface, which is the signal for determine the exotic CPBM phase. (i-l). CPQMC simulation results of pair density structure factor $D_{\mathrm pair}({\bf k})$ in same parameter conditions. Specifically, it shows a rather large peak near $\bf Q$ = ($\pi$, $\pi$), which shows the off-center peak indicate the pairing of fermions with different spins. (m-p). We marked the parameter selection of $n$ and $\alpha$ on 8 $\times$ 8, 12 $\times$ 12, 16 $\times$ 16, 20 $\times$ 20 lattice.  }
    \label{figS1}
\end{figure*}

\section{Phase boundary estimation from the data of finite size system}
This section mainly introduces how to determine various phase regions in the phase diagram from the CPQMC data. In addition to the analysis of $N^{\rm eff}_{pair}$ introduced in the main text, we also defined the density structure factor to further describe the distribution of the Cooper pair in the momentum space. The density structure factor of Cooper pairs is defined as
$D_{\mathrm pair}({\bf k}) = (1/N)\sum_{i,j} \mbox{exp}[i{\bf k}({\bf r}_i-{\bf r}_j), \langle {n}_{bi}  {n}_{bj}  \rangle$, where the Cooper pair number operator is defined as
$n_{bi} = b^{\dagger}_i b_i = n_{i \uparrow} n_{i \downarrow}$, where $b_{i}^\dag = c_{i \uparrow}^\dag c_{i\downarrow}^\dag$ and $n_{i \uparrow}= c_{i \uparrow}^\dag c_{i\uparrow} $.


\begin{figure*} 
    \centering
    \includegraphics[width=0.4\linewidth]{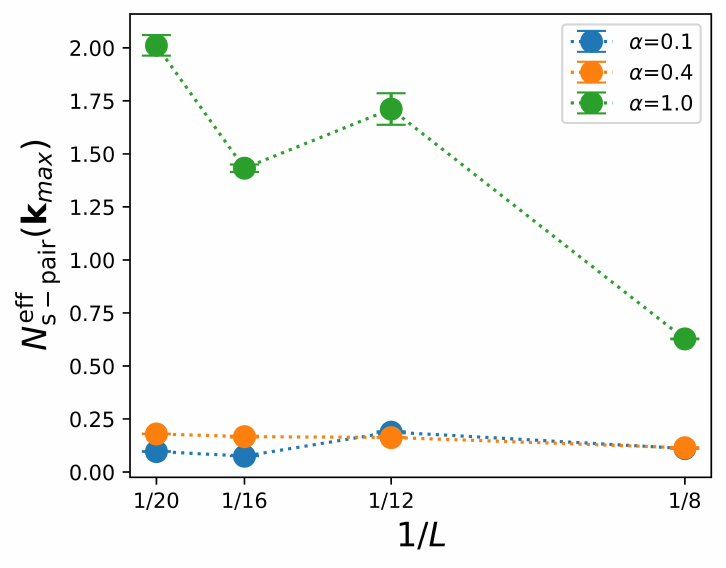}
    \caption{(Color online) Lattice size effect in different phase. CPBM phase for $\alpha =0.10$ (blue), $s$-SF phase for $\alpha =0.40$ (orange), $s$-SF phase for $\alpha = 1.00$ (green) at $n$ $\sim$ 0.63 with $U= -3$ on 8 $\times$ 8, 12 $\times$ 12, 16 $\times$ 16, 20 $\times$ 20 lattice. We can see that in CPBM region, the value of $N^{\rm eff}_{\mathrm s-pair}({\bf k}_{\rm max})$ has always been stable, which is represent that the CPBM phase will always exist in large or even infinite lattice size. As $\alpha$ increases, we can clearly see that the value of $N^{\rm eff}_{\mathrm s-pair}({\bf k}_{\rm max})$ increase, so $s$-SF is slightly suppressed at moderately anisotropic region. When $\alpha = 1.00$, we can clearly see that as the lattice size increases, the value of $N^{\rm eff}_{\mathrm s-pair}({\bf k}_{\rm max})$ is continuously increasing, so $s$-SF is significantly diverge at weak anisotropic region.  }
    \label{figS3}
\end{figure*}


\begin{figure*} 
    \centering
    \includegraphics[width=0.8\linewidth]{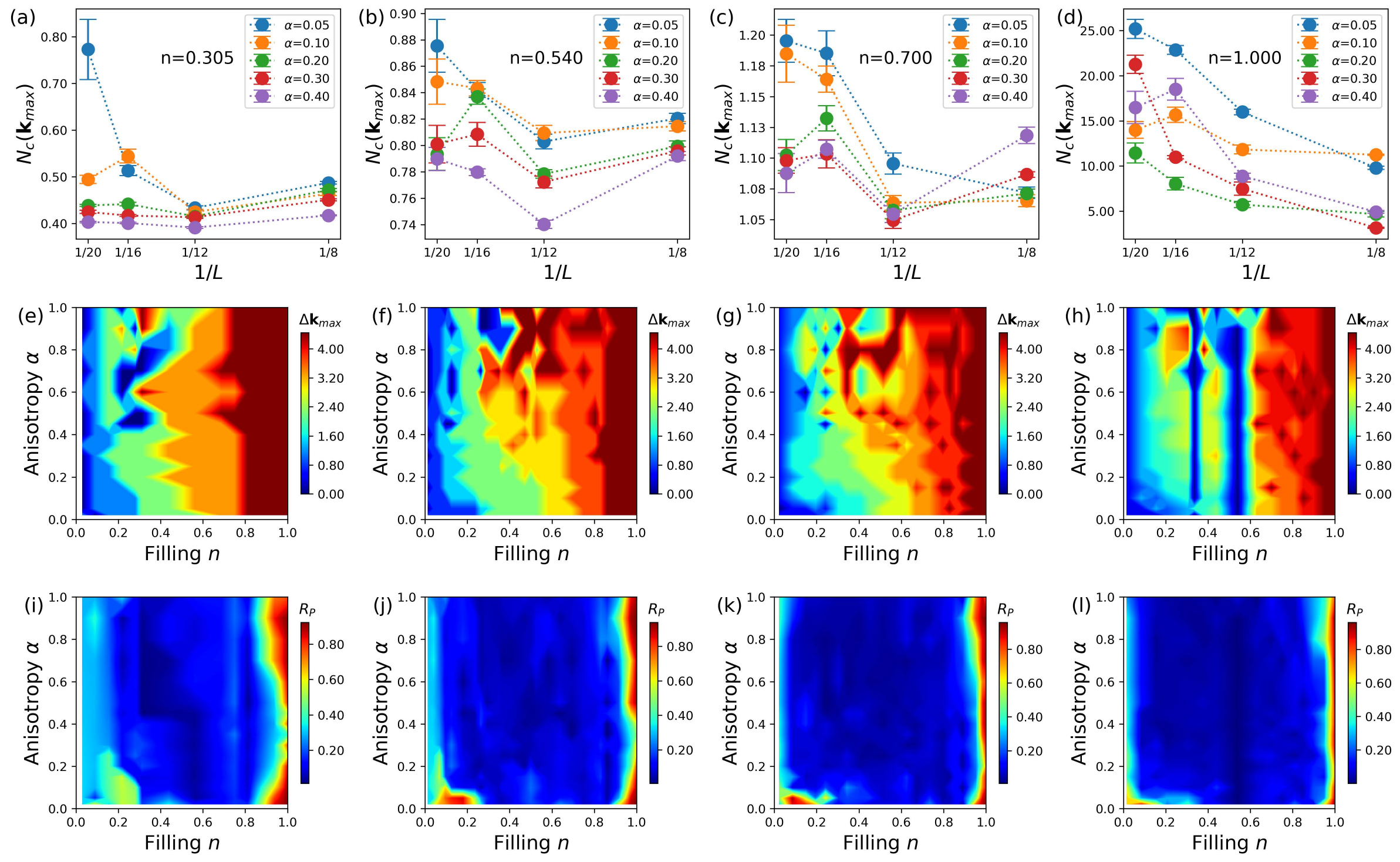}
    \caption{(Color online) Determination of the IDW and CDW. (a-d). Lattice size effect of charge structure factor $N_{\mathrm c}({\bf k})$ for $\alpha$ = 0.05, 0.10, 0.20, 0.30, 0.40 at $U= -3$ on 8 $\times$ 8, 12 $\times$ 12, 16 $\times$ 16, 20 $\times$ 20 lattice for different $n$. $\bf a$, $n$ $\sim$ 0.305, $\bf b$, $n$ $\sim$ 0.54, $\bf c$, $n$ $\sim$ 0.70, $\bf d$, $n$ $\sim$ 1.00. The divergence of lattice effect represents the existence of IDW/CDW phase. (e-h). CPQMC simulation result for distinguishing IDW and CDW regions at $U= -3$ on 8 $\times$ 8, 12 $\times$ 12, 16 $\times$ 16, 20 $\times$ 20 lattice, where $\Delta {\bf k}_{\rm max}=\sqrt{{\bf k}_{\rm max}^2(x)+{\bf k}_{\rm max}^2(y)}$, ${\bf k}_{\rm max}(x)$, ${\bf k}_{\rm max}(y)$ are the coordinates of the maximum value in the charge structure factor. When $\Delta {\bf k}_{\rm max}$ is the maximum value (${\bf k}_{\rm max}(x)={\bf k}_{\rm max}(y)=\pi$), the charge structure factor condensed at $\bf Q$ = ($\pi$, $\pi$) in momentum space, the system display CDW order in phase diagram.  (i-l). CPQMC simulation result for determine the region of IDW and CDW at $U= -3$ on 8 $\times$ 8, 12 $\times$ 12, 16 $\times$ 16, 20 $\times$ 20 lattice, where $R_p=\frac{p-\bar{p}}{p}$, $p$ is the maximum peak value in the charge structure factor and $\bar{p}$ is the average value of the points next to the maximum peak. The blue areas are $R_p\simeq0$, representing the charge structure factor dispersed in momentum space, not forming a peak.If $R_p\gg 0$, it indicates that the IDW condensed at $\bf Q$ = ($2k_F$, $2k_F$) with a sharp peak, so we believe that there exist a IDW phase under this parameter. At small n ($0.05<n<0.2$), a relatively large peak structure appears, but the area becomes smaller as the lattice size increases, which is not a stable state due to the influence of lattice effects. When $n > 0.95$, while CDW order condensed at $\bf Q$ = ($\pi$, $\pi$) displaying a very large peak, the peak intensity may slightly decrease in the regime of intermediate anisotropy $\alpha$.  }
    \label{figS5}
\end{figure*}

We have shown the determination of the CPBM region in Fig.\ \ref{figS1}(a)-(d), on 8 $\times$ 8, 12 $\times$ 12, 16 $\times$ 16, 20 $\times$ 20 lattice, with non-zero peak in $N^{\rm eff}_{s-pair}$. We can see at small lattice size, the distribution of CPBM regimes is relatively scattered, but as the lattice size increases, the distribution area gradually expands. 
Because of constraints in computational resources availability, our calculations were limited to a maximum lattice size of 20 $\times$ 20. From the trend of change, we expect that under infinite lattice points N $\rightarrow\infty$, the CPBM regimes will be connected as a whole, reaching its maximum crtical $\alpha$ at $n$ $\sim$ 0.8, as shown in Fig. 1(a) in the main text. Meanwhile, we draw the Bose surface of features in CPBM phase at different lattice size in Fig.\ \ref{figS1}(e)-(h), selecting on different lattice at $U= -3$ for $\alpha = 0.10$ with $n$ $\sim$ 0.85. As the lattice size increases, the nonzero momentum Bose surface persists and its shape becomes conspicuous. 
We also give the pair density structure factor $D_{\mathrm pair}({\bf k})$ consistent with Fig.\ \ref{figS1}(e)-(h) in Fig.\ \ref{figS1}(i)-(l), which can help us understand the formation of CPBM phase. Specifically, it shows a rather large peak near $\bf Q$ = ($\pi$, $\pi$), which shows the off-center peak indicate the pairing of fermions with different spins. 
Fig.\ \ref{figS3} show the lattice size effect in different anisotropy $\alpha$, which reflects that CPBM is a stable phase.
All of these provided evidence for the existence of CPBM phase and its parameter regions.

IDW along the lattice diagonals exists at strong anisotropy, and having singular features condenses at non zero momentum point $\bf Q$ = ($2k_F$, $2k_F$), gradually diffusing to point $\bf Q$ = ($\pi$, $\pi$) convert to CDW as $n$ approached to half filling.
The boundary points of the IDW region in phase diagram are determined by whether the density waves condensed at $\bf Q$ = ($2k_F$, $2k_F$) point with a sharp peak in momentum space, which we have clearly displayed in Fig.\ \ref{figS5}. From the finite size data in Fig.\ \ref{figS5}(a)-(d), we find the increase of condensation with lattice size in IDW phase. With larger $n$, IDW persists to larger anisotropy up to $\alpha=0.10$. As $n$ gradually increases to half filling ($n$ $>$ 0.95), we combine Fig.\ \ref{figS5}(e)-(h) and Fig.\ \ref{figS5}(i)-(l) to determine a strong CDW order exhibit, which are less sensitive to change of anisotropy. If the density wave condensed at $\bf Q$ = ($\pi$, $\pi$), IDW naturally converts to CDW in our phase diagram.

\section{Numerical methods}

\subsection{Constraint path quantum Monte Carlo~(CPQMC)}

The CPQMC method is a 
quantum Monte Carlo method with a constraint path approximation~\cite{Zhang1995-hn,Zhang1997-mr}. In CPQMC, the ground state wave function $\psi^{(n)}$ is represented by a finite ensemble of Slater determinants, i.e.,  
$\left|\psi^{(n)}\right\rangle \propto \sum_k\left|\phi_k^{(n)}\right\rangle$,
where $k$ is the index of the Slater determinants, and $n$ is the number of iteration. The overall normalization factor of the wave function has been omitted here. The propagation of the Slater determinants dominates the computational time, as follows
\begin{equation}
\left|\phi_k^{(n+1)}\right\rangle \leftarrow \int d \vec{x} P(\vec{x}) B(\vec{x})\left|\phi_k^{(n)}\right\rangle .
\label{propagation}
\end{equation}
where $\vec{x}$ is the auxiliary-field configuration, that we select according to the probability distribution function $P(\vec{x})$. The propagation includes the matrix multiplication of the propagator $B(\vec{x})$ and $\phi_k^{(n)}$. 
After a series of equilibrium steps, the walkers are the Monte Carlo samples of the ground state wave function $\phi^{(0)}$ and ground-state properties can be measured. 

The random walk
formulation suffers from the sign problem because of the fundamental symmetry existing between the fermion ground state $\left|\psi_0\right\rangle$ and its negative $-\left|\psi_0\right\rangle $. 
In more general cases, walkers can cross $\mathcal{N}$ in their propagation by $e^{-\Delta \tau H}$
whose bounding surface $\mathcal{N}$ is defined by $\left\langle\psi_0 \mid \phi\right\rangle=0$ and is in general $unknown$.
Once a random walker reaches $\mathcal{N}$, it will make no further contribution to the representation of the ground state since
\begin{equation}
\left\langle\psi_0 \mid \phi\right\rangle=0 \Rightarrow\left\langle\psi_0\left|e^{-\tau H}\right| \phi\right\rangle=0 \quad \text { for any } \tau .
\label{no contribution}
\end{equation}

Paths that result from such a walker have equal probability of being in either half of the Slater-determinant space. Computed analytically, they would cancel, but without any knowledge of $\mathcal{N}$, they continue to be sampled in the random walk and become Monte Carlo noise. 

The decay of the signal-to-noise ratio, i.e., the decay of the average sign of $\left\langle\psi_T \mid \phi\right\rangle$, occurs at an exponential rate with imaginary time.
To eliminate the decay of the signal-to-noise ratio, we impose the constrained path approximation. It requires that each random walker at each step has a positive overlap with the trial wave function $\left|\psi_T\right\rangle$ :

\begin{equation}
\left\langle\psi_T \mid \phi_k^{(n)}\right\rangle>0 .
\label{constrained path approximation}
\end{equation}

This yields an approximate solution to the ground-state wave function, $\left|\psi_0^c\right\rangle=\Sigma_\phi|\phi\rangle$, in which all Slater determinants $|\phi\rangle$ satisfy Eq.~\ref{no contribution}. From Eq.~\ref{constrained path approximation}, it follows that this approximation becomes exact for an exact trial wave function $\left|\psi_T\right\rangle=\left|\psi_0\right\rangle $.

\begin{figure*}[htb!]
    \centering
    \includegraphics[width=1\linewidth]{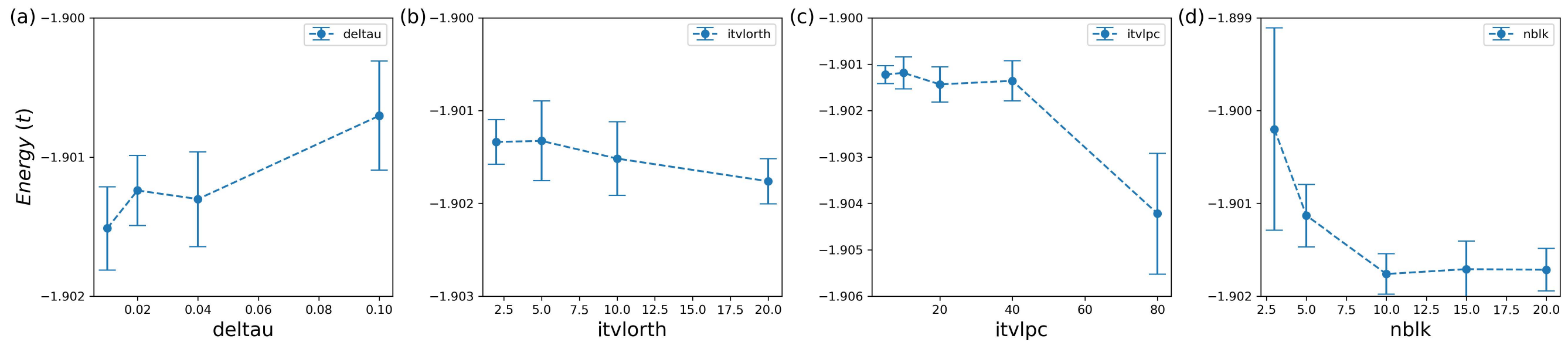}
    \caption{(Color online) Choice of hyperparameters in CPQMC.  (a-d). We shows the convergence of the total energy per particles with change of hyper-parameters used in CPQMC calculations for $\alpha =0.10$ at $U= -3$ with $n=$ 0.85 on 20 $\times$ 20 lattice. The parameters we ultimately chose are as follows: the Trotter step size is deltau $=0.02$, the number of orthonormalization interval is itvlorth $=10$, the Pop control interval in measurement phase is itvlpc $=20$, the number of blocks for relaxation is nblk $=10$. Some other parameters used in CPQMC calculations are provided in the method.   }
    \label{figS9}
\end{figure*}

Next, we will introduce the parameter selection problem we use in the CPQMC calculation process.
We choose the lattice size as 8 $\times$ 8, 12 $\times$ 12, 16 $\times$ 16, 20 $\times$ 20, so the system size $N$ are 64, 144, 256, 400 respectively. Meanwhile, the filling $n$ = $(n_{\uparrow}+n_{\downarrow})/N$, and we need to ensure that the number of ${\uparrow}$ and ${\downarrow}$ electrons is the same and that there cannot be a degenerate state, so $n$ only can take some fixed values with compete shells for a lattice size. We select $n$ is taken at intervals of 0-1 based on available compete shells, $\alpha$ is taken as 0.02, 0.05, 0.10, 0.15, 0.20, 0.25, 0.30, 0.35, 0.40, 0.45, 0.50, 0.60, 0.70, 0.80, 0.90, 1.00 (because the system varies more plentiful when $\alpha$ is small), and $U$ = -3 (the most extensive region of CPBM phase). We have marked the parameter selection of $n$ and $\alpha$ corresponding to different lattice size in Fig.\ \ref{figS1}(m)-(p).
The details of hyper-parameters in CPQMC calculations are given as follows. The number of walkers is 1000, the number of blocks for relaxation is 10, the number of blocks for growth estimate is 3, the number of blocks after relaxation is 10, the number of steps in each block is 320, the Trotter step size is 0.02, the growth-control energy estimate is -50. The number of population control interval in relaxation phase is 10, the number of population control interval in measurement phase is 20, the number of orthonormalization interval is 10, the measurement interval is 40, the back propagation steps is 40. We provide the stability of the total energy per particles in CPQMC calculations with certain parameter changes (Fig.\ \ref{figS9}).
As the lattice size increases, the computational time increases quickly, with the computational complexity dominated by the matrix multiplication. When dealing with large lattice, we design a parallelization scheme to accelearate the simulation. A convenient way to parallelize the CPQMC program is to distribute the random walkers and its associated information evenly to different cores. Three major phases are involved in the simulation, (1) relaxation, (2) growth estimation, and (3) measurement. We do parallel computing the phases of relaxation and measurement, where the majority of the computational time located. In growth estimation, the iterative update of weight leads to frequent communication between threads, and complication of parallel scheme, so we omit its parallelization. 

\subsection{Functional renormlaization group method~(FRG)}

The idea of FRG \cite{Wetterich1991-gt} is to obtain the one-particle-irreducible 4-point interaction vertices $\Gamma_{1234}$ (where numerical index labels single-particle state) for quasi-particles above a running infrared energy cut off $\Lambda$ (which we take as the lower limit of the Matsubara frequency). Starting from $\Lambda=\infty$ where $\Gamma$ is specified by on-site interaction $U$, the contribution to the flow (toward decreasing $\Lambda$) of the vertex is given by 

\begin{equation}
\partial \Gamma / \partial \Lambda = -1/2 {\rm P} \partial \chi_{pp} /\partial \Lambda {\rm P} + {\rm C} \partial \chi_{ph} / \partial \Lambda {\rm C} - {\rm D} \partial \chi_{ph} / \partial \Lambda {\rm D}\\
\end{equation}
Where, P, C, D are rearrangements of $\Gamma$ in the pairing(P), crossing (C) and direct(D) channels, and $\chi_{pp/ph}$ are corresponding susceptibilities in the particle-particle (PP) and particle-hole (PH) channels.

At each stage of the flow, we decompose $\Gamma$ in terms of eigen scattering modes (separately) in the PP (pairing) and PH (charge density and spin density) channels in the Brillouin Zone to find the negative leading eigenvalues (NLE) $S$ ($S<0$), with the associated eigenfunctions describe the internal microscopic structure of the modes. The decrease of NLE $S$, with the flow of $\Lambda$, indicates the enhancement of fluctuation in the corresponding channel, and the divergence of which signals an emerging order at the associated scattering momentum. In the paper, we find that the dominant NLE in the PH channel is always charge density wave in the phase diagram, except the associated scattering momentum changed. However, the dominant NLE in PP channel include non-zero momentum pairing and $s$-SF. More technical details of FRG can be found elsewhere \cite{Wang2012-uu,Wang2019-at}.  

We used two complementary numerical techniques to study the 2D system with spin-dependent anisotropy, CPQMC and FRG. While the result of CPQMC provides physical observables like correlations that is important in defining Bose metal, the result is under controllable approximation in the presence of sign problem. The FRG is a highly reliable method in term of both algorithmic foundation and the achievable lattice size, but its judgement would be based on divergence of eigenvalue on the quantity that defining known phase. In both we find consistent signals of Bose surface in CPBM phase.

 

\bibliography{ref}